# Absence of Tunneling Character in c-axis Transport of SmFeAsO$_{0.85}$ Single Crystals

**Jae-Hyun Park, Hyun-Sook Lee, and Hu-Jong Lee**[*]

*Department of physics, Pohang University of Science and Technology, Pohang 790-784, Republic of Korea*

**B. K. Cho**

*Materials Science and Engineering, Gwangju Institute of Science and Technology, Gwangju 500-712, Republic of Korea*

**Yong-Joo Doh**

*Department of Display and Semiconductor Physics, Korea University Sejong Campus, Chungnam 339-700, Republic of Korea*

We made electrical transport measurements along the *c*-axis of SmFeAsO$_{0.85}$ single crystals, in both three- and four-terminal configurations, focusing on examining the possible formation of Josephson coupling between FeAs superconducting layers. Anisotropic bulk superconductivity was observed along the *c*-axis, but without Josephson coupling, as confirmed by the absence of both the hysteresis in current-voltage curves and the modulation of the critical current by the in-plane magnetic fields. The variation of the critical currents for different magnetic-field directions gives the anisotropy ratio of γ ~ 5. This three-dimensional electronic structure of our iron pnictide superconductors is in clear contrast to the two-dimensional one observed in the cuprate superconductors, another stacked superconducting system.

PACS number:

[*]Email: hjlee@postech.ac.kr

Fax: +82-54-279-5564

# 1. Introduction

In recent years, the discovery of superconductivity in iron pnictides [1] has attracted high interests while consensus is yet to be reached on the detailed character of the superconductivity in the material. It is expected that studies on the iron pnictide superconductors may provide a fresh insight into clarifying the origin of the cuprate superconductivity. As in the cuprate superconductors, the superconducting phase in the iron pnictides emerges upon doping a parent antiferromagnet with electrons or holes, and the superconducting transition temperature ($T_c$) reveals a dome-shape dependence on the doping concentration. Similar to cuprates, iron pnictide superconductors have layered structures of alternating Fe-Pn superconducting layer and doped charge-reservoir layer (RE-O layer for REFePnO superconductors and AE layer for $AEFe_2Pn_2$, where RE, AE and Pn indicate the rare-earth, alkaline-earth, and pnictogen elements, respectively) along the $c$-axis [2]. A natural Josephson coupling is established between the superconducting layers in highly anisotropic cuprate superconductors [3]. Thus, examining the anisotropic transport properties along the $c$-axis to explore the possible establishment of Josephson coupling in iron pnictides with layered structures is of high interest. The formation of natural Josephson junctions is considered to be very helpful in clarifying the superconducting pairing symmetry [4]. It is also crucial for applications to devices such as SQUIDs [5], Josephson voltage standards [6], and THz oscillators [7].

Earlier studies on the parent materials of $AEFe_2Pn_2$ superconductors [8-9] such as $SrFe_2As_2$ and $BaFe_2As_2$ revealed a relatively large anisotropy ratio of $\gamma_\rho = \rho_c/\rho_{ab} > 100$. In addition, the band-structure calculations for parent materials of REFePnO superconductors suggest a two-dimensional electronic structure [10]. Recent measurements of the upper critical field, $H_{c2}$, of REFePnO superconductors show a larger anisotropy ratio ($\gamma_H = H_{c2}^{ab}/H_{c2}^{c}$) than AEFe$_2$Pn$_2$ family [11-13]. This is consistent with that the spacing between adjacent Fe-As layers in REFePnO superconductors is larger than in AEFe$_2$Pn$_2$ superconductors [12]. On the other hand, studies on the REFePnO superconductors, in particular, on anisotropic properties of the materials using transport measurements, are impeded by the difficulties of synthesizing high-quality single crystals of convenient size for sample fabrication. With difficulties of thinning the crystals by successive cleaving, putting contact leads on thick crystals poses challenging difficulties, often leading to disconnection of leads at the crystal edge. Thus, most of the effort for investigating anisotropic transport properties in the iron pnictides has been focused on single crystals of AEFe$_2$Pn$_2$ families.

Fluorine-free SmFeAsO$_{1-x}$ single crystals, grown in a high-pressure furnace in our previous study [14], exhibit relatively high anisotropy ratio among the iron pnictide superconductors. Here, we report on the *c*-axis transport properties for various temperatures and magnetic fields for different field angles to examine the possible establishment of the interlayer Josephson coupling and the anisotropicity in transport properties of SmFeAsO$_{0.85}$. Difficulties in attaching measurement leads on thick single crystals with small lateral size were overcome by adopting the electron-beam patterning, planarization with thick insulating layer, and SiN stencil-masking techniques for the deposition of contact leads. For a specimen of highly anisotropic material, precise electrical transport properties of the normal or the superconducting state are obtained in specific measurement configurations of the specimen. In this study, comprehensive electrical transport measurements on SmFeAsO$_{0.85}$ single crystals were made by adopting three- and four-terminal measurement configurations to have more conclusive insights into the *c*-axis transport of the material. Temperature dependence of the resistance, current-voltage characteristics, and the magnetic field dependence of the critical currents, along the *c*-axis, reveal weakly anisotropic bulk transport properties of the material rather than formation of the Josephson weak links. This result is consistent with our recent report on the anisotropic behavior of $H_{c2}$ in SmFeAsO$_{0.85}$ single crystals [12].

## 2. Experimental procedure

Fluorine-free SmFeAsO$_{0.85}$ single crystals were synthesized by adopting the self-flux method under high-temperature (1350 – 1450 ℃) and high-pressure (~ 3.3 GPa) conditions with stoichiometric starting materials of SmAs, Fe$_2$O$_3$, and Fe. Details of the single-crystal growth are described in our previous report [14]. Plate-like single crystals with lateral size of 7 – 10 μm and thickness of 0.5 – 2 μm were attached onto sapphire substrates using negative photoresist. After thin Au film (~50 nm) was thermally deposited on a crystal, a mesa structure was formed by using electron-beam lithography and Ar-ion etching [15]. The height of the mesas was determined by the etching rate of 5 nm/min, calibrated by the atomic force microscopic (AFM) measurements, for different total etching time with the beam current and voltage of 20 mA and 300 V, respectively. A mesa was planarized by spinning another layer of negative electron-beam resist up to 500 nm in its thickness, which was patterned and used as an insulating layer for the electrical connections to the top surface of the mesa structure. Then, 100 nm-thick Au electrodes were formed using a stencil mask of SiN, which replaced the multiple processes in arranging electrodes by depositing an Au layer, electron-beam lithography, and subsequent Ar-ion etching. A mesa-structured device D1 [the lower-left configuration of Fig. 1(a)] was fabricated in a three-terminal measurement configuration by adopting a SiN mask.

We also made measurements on another mesa D3 in a four-terminal configuration [the lower-right configuration of Fig. 1(a)], fabricated by adopting the method described in Ref. [15]. The optical image of the device D1 is shown in Fig. 1 (a), along with the schematics of the three- (lower left) and four-terminal (lower right) measurement configurations. Table 1 shows the details of the physical parameters of the devices used. Resistance vs temperature ($R$–$T$) and current-voltage ($I$–$V$) curves for various temperatures and external magnetic fields were taken in the current-bias mode in both three- and four-terminal configurations. Magnetic-field tilt-angle dependence of the $c$-axis magnetoresistance was also taken *in-situ* by rotating the sample stage inside a superconducting solenoid by a stepper motor located at room temperature, with the angle-rotation resolution of $0.1°$.

## 3. Results and discussion

Resistivity along the $c$-axis ($\rho_c$) of the device D1 was measured in the temperature range of $T = 4 - 280$ K in zero external magnetic field for a constant bias current of $I = 1$ µA [see Fig. 1(b)]. Similar $\rho_c$–$T$ curve obtained from the device D3 is shown in the inset. The superconducting transition temperatures ($T_c$'s) (obtained by the criterion of 50 % of the normal-state resistance $R_n$), for the devices D1 and D3, are about 53.8 K and 51.9 K, respectively, along with the corresponding transition widths, $\Delta T_c$'s, (by adopting the criterion of $10 - 90\%$ of $R_n$) of about 0.7 K and 2.3 K. The discrepancy of $T_c$ between the devices D1 and D3, fabricated out of crystals of the same batch, is presumably due to local variations of the oxygen vacancies in SmFeAsO$_{1-x}$ single crystals. The value of $\rho_c$ obtained from the resistance of the mesa geometry at the temperature for 90% of $R_n$ is about 5 mΩ·cm for the device D3 in four-terminal configuration, which is about 5 times smaller than 25 mΩ·cm [obtained by subtracting the contact resistance of ~40 mΩ·cm; refer to Fig. 1(b)] for the device D1 in three-terminal configuration. The reduction of the normal-state resistance in D3, however, may have been caused by the nonuniformity of the injected current over the junction area in the four-terminal configuration. Higher uniformity of the current flow in the junction for D1 in a three-terminal configuration leads to more accurate normal-state transport properties. The value of $\rho_c$, measured from the device D1, is three orders of magnitude larger than the $ab$-plane resistivity $\rho_{ab} \sim 0.08$ m$\Omega\cdot$cm for our SmFeAsO$_{0.85}$ single crystals [14]. Monotonous decrease of $\rho_c$ with decreasing temperature above $T_c$ suggests the metallic characteristics of the $c$-axis transport of our crystals rather than the semiconducting behavior, which is generally observed in the $c$-axis transport of Bi$_2$Sr$_2$CaCu$_2$O$_{8+x}$ (Bi-2212) single crystals, except for overdoped regime [16]. The increase of the interfacial resistance of the device D1 with decreasing temperature below $T_c$ is caused by the decrease of the conductance near zero bias due to the finite barrier formed at the interface between the top of the mesa and the normal-metallic electrodes [17-18].

Temperature-dependent $I$–$V$ curves are shown in Fig. 2 for two different devices, the dimensions and the respective measurement configurations of which are listed in Table 1. For the $I$–$V$ curves in Figs. 2 (a), the linear contact resistance (about 5 $\Omega$ for D1, with little temperature dependence) was subtracted numerically. Irrespective of the measurement configurations, the temperature dependence of $I$–$V$ curves turns out to be similar with a steady increase of $I_c$ with lowering temperatures, except for the asymmetry of the negative-bias branches in Fig. 2(b) for the device D3. The asymmetry is mainly attributed to the non-uniform current flow in the junction area. The asymmetry is also partly caused by the hysteresis [see the inset of Fig. 2(a)] in the voltage response to a current bias swept from a positive to a negative polarity direction. Joule heating, from the interfacial resistance between Au electrodes and the top of the mesa structure, may have caused the hysteresis for a bias current above $I_c$. In the inset of Fig. 2(a), the discrepancy of the critical currents $I_c$, due to the hysteresis is about 0.3 mA, which corresponds to the temperature difference of 10 K according to the temperature dependence of $I_c$ near $T = 10$ K. This increase of the temperature can be caused by the maximally used self-heating of ~1 mW in the device D1 [see the inset of Fig. 2(a)], as observed in the interlayer tunneling spectroscopy for Bi-2212 single crystals [19]. Multiple quasiparticle branches, which is observed in the $c$-axis transport measurements for the intrinsic Josephson junctions, is absent in the $I$–$V$ curves. The value of $I_c R_n$ per junction ($I_c R_n /N$; $N$ is the number of Fe-As layers along the $c$-axis) is 0.36mV for the device D1. These values are much smaller than the zero-temperature superconducting gap energies, $\Delta_1(0) = 6$ meV and $\Delta_2(0) = 18$ meV, observed in the Andreev-reflection spectroscopy for $SmFeAsO_{0.8}F_{0.2}$ [18]. This implies that the observed values of $I_c R_n$ do not stem from the formation of intrinsic tunneling junctions. Thus, no hysteretic multiple quasiparticle branches are observed in our experimental range for both measurement configurations.

Temperature dependence of $I_c$ is very useful to examining the possible formation of the Josephson coupling along the $c$-axis. $I_c$ for a given temperature was determined by averaging the absolute values of positive and negative critical currents, which were taken for the $c$-axis voltage corresponding to $V = \pm 0.1$ mV. The critical current density along the $c$-axis ($J_c$) at the base temperature is in the range of $1.6 \times 10^5$ – $1.8 \times 10^5$ A/cm$^2$ for our devices. The temperature dependences of the normalized critical current ($I_c/I_{c0}$) obtained below $T_c$ is shown in the inset of Fig. 2 (b) as a function of the reduced temperature $T/T_c$, which is not sensitive to the specific measurement configurations. With lowering temperature, the normalized $I_c$ exhibits a linear variation near $T_c$, then gradually turns into a slight downturn curvature, and finally saturates into the maximum value $I_{c0}$ at the lowest temperatures. Similar temperature dependence can be obtained from the Ambegaokar-Baratoff (A-B) relation [20], $I_c=(\pi\Delta/2eR_n)\tanh(\Delta/2k_BT)$, as indicated by the solid line in the inset of Fig. 2(b), which is a theoretical expectation for the temperature dependence of $I_c$ of a superconductor-insulator-superconductor (SIS) Josephson junction. The value of $eI_{c0}R_{n,single}/\pi\Delta_1(0)$, however, is about 0.02 for the device D1, which is lower than the expected value of A-B relation, $eI_{c0}R_{n,single}/\pi\Delta_0 = 0.5$, where $R_{n,single}(=R_n/N)$ is the normal-state resistance of a single junction and $\Delta_0$ is the zero-temperature gap energy [21]. Consequently, in spite of the qualitative resemblance, the temperature dependence of $I_c$ does not follow the tunneling Josephson junction behavior. The possibility of the formation of superconductor–normal-metal–superconductor (SNS) Josephson junction along the $c$-axis is also excluded by the absence of exponential temperature dependences of $I_c$ of the form $I_c \propto \exp(-L/L_T)$, where $L_T$ $(=\sqrt{\hbar D/2\pi k_B T}$ ; $\hbar$ is the Planck's constant, $D$ the diffusivity, $k_B$ the Boltzmann constant) is the characteristic thermal length in the diffusive limit and $L$ is the length of the junction [22] .

Another representative characteristic of the intrinsic Josephson coupling can be obtained from the magnetic-field modulation of $I_c$ [23]. Figure 3 (a) depict $I$–$V$ curves for the device D1 in an in-plane external magnetic field ranging from 0 T to 6 T at $T = 50$ K. Similar measurements were made at several different temperatures. Magnetic field dependence of $I_c$ is shown in Fig. 3(b) for in-plane magnetic field direction for five different temperatures. A diffraction-like "Fraunhofer" pattern is expected in a typical Josephson junction in a magnetic field applied in parallel with the $ab$-plane with the magnetic field periodicity of $H = \Phi_0/cL = 1$ T, corresponding to one flux quantum enclosed per junction. Here, $\Phi_0$ is the flux quantum, $c \sim 0.84$ nm is the $c$-axis lattice constant, and $L = 2$ μm is the lateral dimension of the mesa structure. No symptom of oscillatory variations of $I_c$, as expected for a tunneling Jospehson junction, is observed in our experimental range. The device D3 also gives similar result of monotonous decrease of $I_c$ with the external magnetic field (not shown here). Moreover, as shown in inset of Fig. 3 (a), the magnetoresistance in the current bias near $I_c$ for device D3, with continuous sweeping of magnetic fields, exhibits monotonous behavior rather than oscillatory modulation.

For a Josephson junction in a long junction limit, where the lateral dimension of a Josephson junction is much longer than the Josephson penetration depth $\lambda_J$, $I_c$ can monotonously decrease, with $H$ without field modulation, due to the spatial distribution of the superconducting phase difference in the $ab$-plane or the self-field effect. On the other hand, in our devices, $\lambda_J$ is expected to be in the range of 3–10 nm, estimated from the measurement of the anisotropy ratio of London penetration depth [24], $\lambda_L$, based on the relation [25] of $\lambda_J = (\Phi_0 d/2\pi\mu_0\lambda_{L,ab}{}^2 J_c)^{1/2} = \lambda_{L,c}/\lambda_{L,ab}((t+d)d)^{1/2}$. Here, $d$ is the thickness of the superconducting layer, $t$ is the thickness of the insulating layer, $\mu_0$ is the permeability of free space, $\lambda_{L,ab}$ and $\lambda_{L,c}$ are the London penetration depth in parallel with the $ab$-plane and along the $c$-axis, respectively, and $J_c$ is the critical current density along the $c$-axis. This value of $\lambda_J$ is unusually small because of the large $c$-axis critical current density of our devices. In a long-junction limit ($L \gg \lambda_J$), the relation $1-I_c/I_{c0} = (H/H_0)^{1/2}$ is applicable [26-27]. Thus, the $I_c$–$H$ curves for various temperatures are normalized by $1-I_c/I_{c0}$ and are depicted with the horizontal axis for $H^{1/2}$ as shown in Fig. 3(c). The normalized $I_c$'s are averaged for positive and negative polarity of the magnetic fields. In Fig. 3(c), the normalized $I_c$'s clearly deviate from the relation to $H^{1/2}$. If the data at 35 K in Fig. 3(c) are fit to the relation $1-I_c/I_{c0} = (H/H_0)^{1/2}$, the best-fit value of $H_0$ turns out to be 2200 T, which is unphysically high compared to $H_{c2}$ of ReFePnO superconductors [12, 28]. Thus, the in-plane field dependence of the $c$-axis resistance negates the formation of the Josephson tunnel junctions.

Figs. 4 (a) and (b), respectively, show the magnetic field dependence of the $I$–$V$ curves of the device D1 at $T = 50$ K and the $I_c$–$H$ curves with five different temperatures in magnetic fields applied perpendicular to the $ab$-plane. The magnetic field dependence of each figure is similar to the corresponding in-plane field dependence shown in Figs.

3(a) and (b). Although the variation of the *I–V* curves and their $I_c$'s are more rapid than the case of in-plane magnetic fields, the trend of suppression of $I_c$'s resembles each other. In Fig. 4 (c), the normalized $I_c$–*H* curves for both in-plane and out-of-plane magnetic field are shown where $1$-$I_c/I_{c0}$ is plotted as a function of the field strength *H*. As shown in the figure, the curves for different magnetic field orientations but with the same magnetic field strength well coincide with each other, where the scale of the right vertical axis for the in-plane magnetic field is reduced by ~ 5 times. The magnetic field dependence of higher sensitivity for $I_c$ along the *ab*-plane is understood by the larger diamagnetization energy for the in-plane field in comparison with the magnetic field perpendicular to the *ab*-plane. All the magnetic-field dependences of the resistance and the critical current along the *c*-axis described above are attributed to the field-induced suppression of the bulk superconducting order, which is reflected in the upper critical field $H_{c2}$. But, it is not due to the existence of the Josephson coupling between superconducting layers and resulting field-induced weakening of the coupling strength along the *c*-axis in the material. Therefore, the scaling factor of ~ 5 for the reduction of $I_c$'s in varying magnetic field strength for the two different field directions implies the anisotropy ratio of $\gamma$ ~ 5 in the material, which stems from the anisotropicity of $H_{c2}$ [12].

## 4. Conclusion

Our *c*-axis transport measurements on SmFeAsO$_{0.85}$ single crystals reveal weak anisotropy without any clue of the existence of the intrinsic Josephson junctions. The coherence length of SmFeAsO$_{0.85}$ single crystals along the *c*-axis [12], $\xi_c$ ~ 0.36 nm inferred from the measurement of $H_{c2}$, is comparable to the lattice constant of SmFeAsO$_{0.85}$ single crystals along the *c*-axis [14], *c* ~ 0.84 nm, as opposed to more than 10 times smaller coherence length than the lattice constant along the *c*-axis in Bi-2212 single crystals [3], where intrinsic Josephson coupling exists. It implies that the *c*-axis transport in our SmFeAsO$_{0.85}$ single crystals reveals the bulk transport properties rather than tunneling transport characteristics. Due to the comparability of the magnitude of two characteristic lengths, the superconductivity of Fe-As layers sandwiching a Sm-O layer diffuses into the whole region of the non-superconducting Sm-O layer. Because of the strong induction of superconductivity in Sm-O layers, SmFeAsO$_{0.85}$ single crystals can be treated as nearly homogeneous materials. The spectroscopic evidence for the three-dimensional electronic structure of iron pnictide superconductors [29-30] rather than two dimensionality of the cuprate superconductors is also supported by the weak anisotropic properties, originating from the corrugation of the Fermi surface along the k$_z$-direction. Therefore, the electrical transport properties along the *c*-axis can be explained by adopting the concept of barely modulating superconducting order parameters along the *c*-axis, which in turn brings about the weak anisotropic superconducting properties of SmFeAsO$_{0.85}$ single crystals with the anisotropy ratio of $\gamma$ ~ 5.

## Acknowledgments

This work was supported by the National Research Foundation through Acceleration Research Grant R17-2008-007-01001-0 and also by POSCO.

**FIGURE CAPTIONS**

Table 1: Geometric dimensions, measurement configurations, and superconducting parameters of SmFeAsO$_{0.85}$ single crystals obtained from the measurements of the $c$-axis transport for the devices D1 and D3. $T_c$ is determined as midpoint of the superconducting transition and $\rho_c$ is calculated from the resistance drop over the transition and geometric dimensions. $J_{c0}$ is the critical current density at the lowest measurement temperature.

FIG. 1. (a) The optical micrograph of the device D1, and schematic view of three- and four-terminal measurement configurations. (b) $\rho_c$–$T$ curve along the $c$-axis of SmFeAsO$_{0.85}$ single crystal for the device D1 in the absence of external magnetic field. The inset shows the expanded view of $\rho_c$–$T$ curves for the device D3.

FIG. 2. $I$–$V$ curves along the $c$-axis measured for (a) the device D1 for $T$ = 10, 25, 30, 35, 40, 45, 47.5, 50, 52, 53, and 55 K, and (b) the device D3 for T = 6.5, 20, 30, 37.5, 42.5, 45, 47.5, 49, 51, 52.5, and 55 K. Inset of (a) depicts the raw $I$–$V$ curve (including the contact resistance) of the device D1 at $T$ = 10 K, exhibiting the hysteresis. Inset of (b) shows the normalized critical currents ($I_c/I_{c0}$) as a function of the reduced temperature ($T/T_c$) for the two different devices as in (a) and (b). The solid line is the A-B relation of SIS junctions.

FIG. 3. (a) $I$–$V$ curves at 50 K for in-plane fields of $H$ = 0, 1, 2.5, 4, 5, and 6 T. Inset: the measurement of the magnetoresistance for the device D3 in the bias current of $I$ = 1 mA at $T$ = 51 K for continuous sweeping of magnetic fields. (b) $I_c$–$H$ curves for different temperatures in in-plane magnetic fields. Lines are guides to eyes. (c) Normalized $I_c$ is shown as a function of $H^{1/2}$ for different temperatures in in-plane magnetic fields. The dashed lines are guides to eyes.

FIG.4. (a) *I–V* curves at 50 K for out-of-plane fields of *H* = 0, 0.5, 1, 1.5, 2, 3, 4, 5, and 6 T. (b) *I_c–H* curves for different temperatures in out-of-plane magnetic fields. (c) Normalized *I_c–H* curves for out-of-plane (left vertical axis) and in-plane (right vertical axis) magnetic fields. Lines connecting data points are guides to eyes. The horizontal dashed line denotes the maximum value for the normalized *I_c–H* curves for out-of-plane magnetic fields.

Table 1

| device | configuration | area ($\mu m^2$) | height (nm) | $T_c$ (K) | $\rho_c$ (m$\Omega \cdot$cm) | $J_{c0}$ ($10^5$ A/cm$^2$) |
|--------|---------------|------------------|-------------|-----------|------------------------------|----------------------------|
| D1 | 3-terminal | $2 \times 2$ | 50 | 53.8 | 25 | 1.8 |
| D3 | 4-terminal | $4 \times 1$ | 40 | 51.9 | 5 | 1.6 |

Fig. 1   (a), (b)

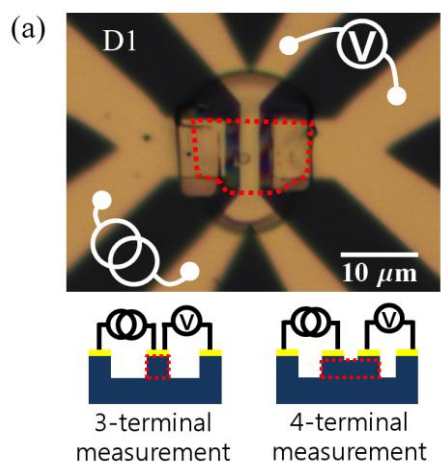

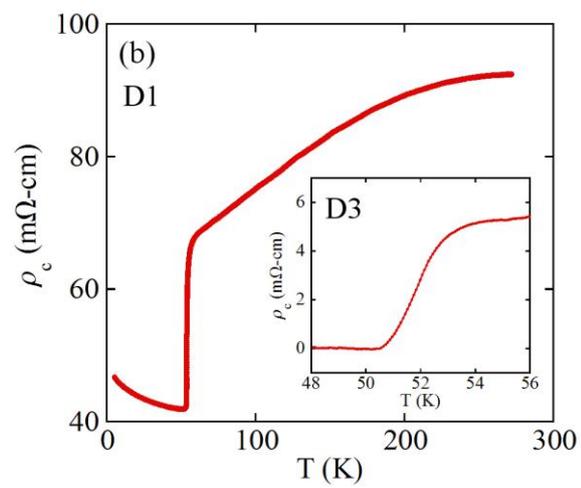

Fig. 2 (a), (b)

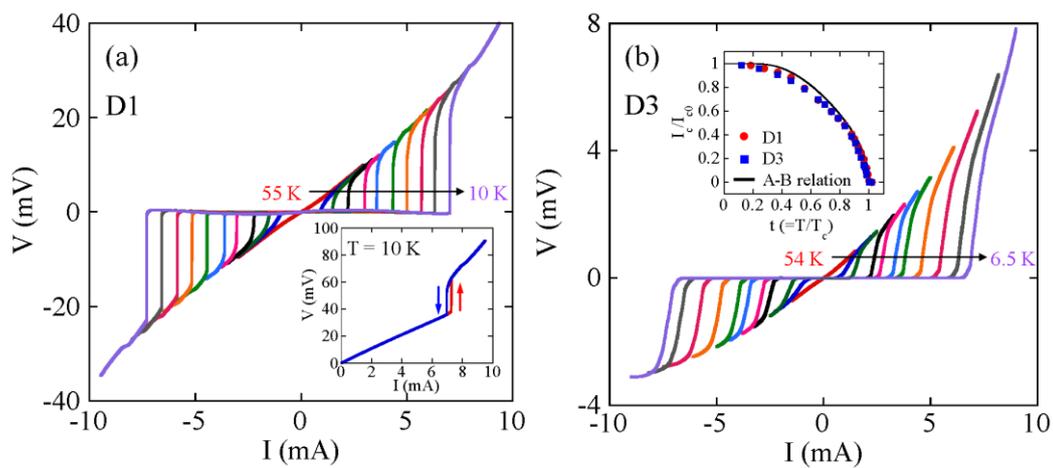

Fig. 3 (a), (b), (c)

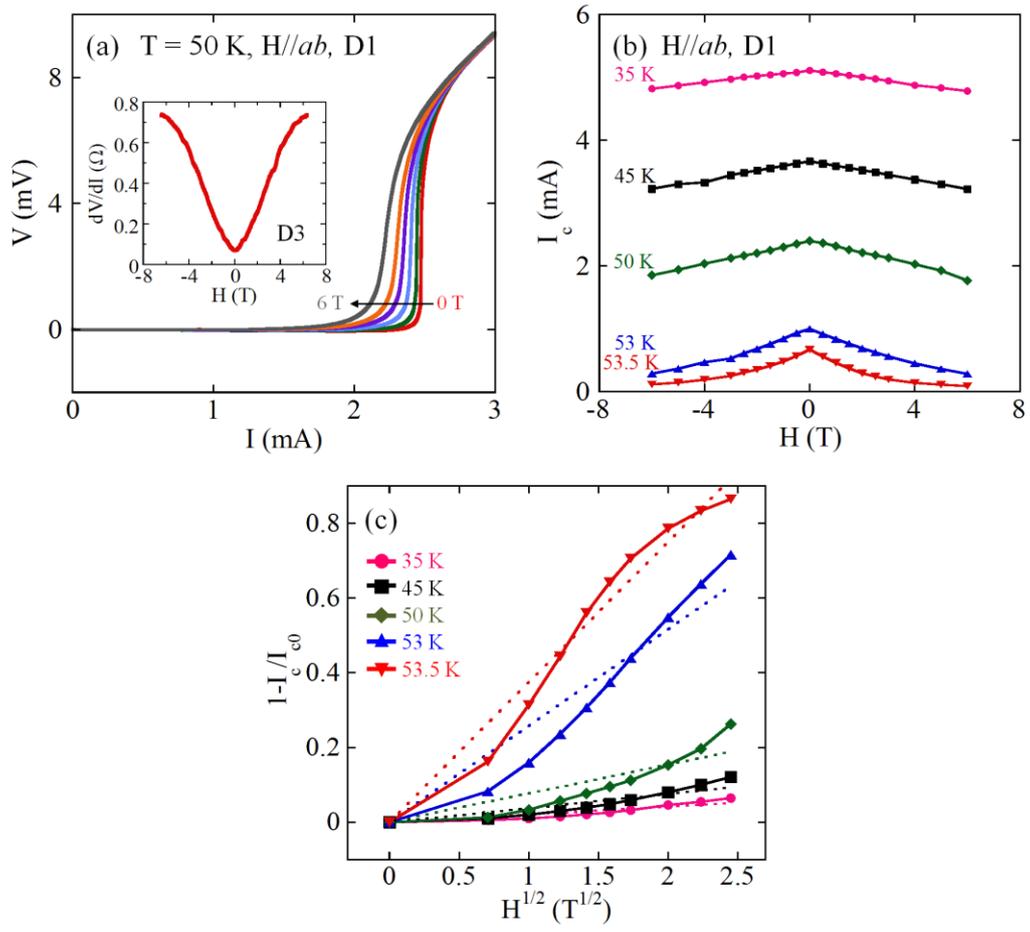

Fig. 4 (a), (b), (c)

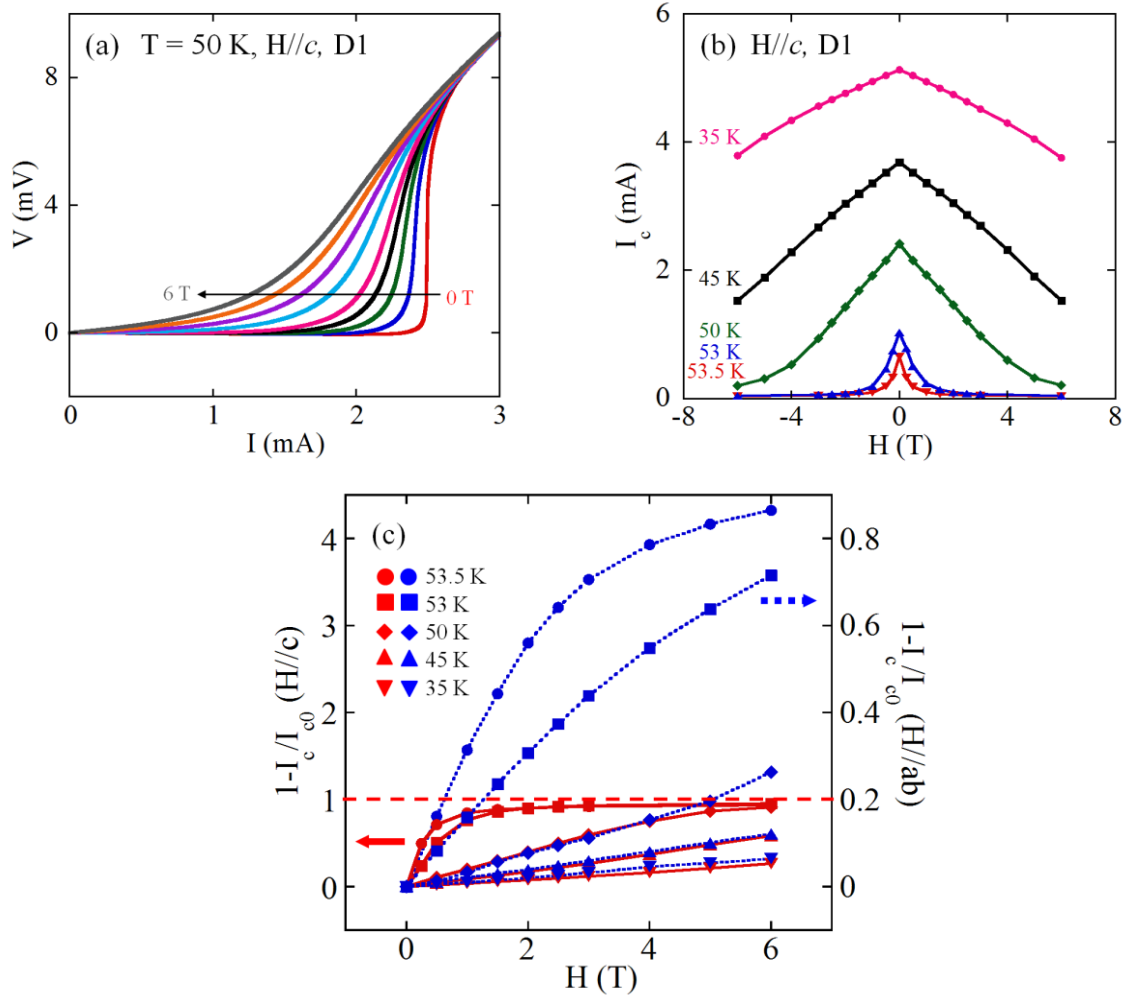